\documentclass[submission, Phys]{SciPost}

\usepackage{amsmath,amsthm,amssymb}
\newcommand{\nc}{\newcommand}
\nc{\rnc}{\renewcommand}
\nc{\nn}{\nonumber}
\nc{\der}{{\partial}}
\rnc{\Im}{{\textrm{Im}\,}}
\rnc{\Re}{{\textrm{Re}\,}}
\nc{\db}{\displaybreak[0]\\}
\nc{\bra}{\langle}
\nc{\ket}{\rangle}
\nc{\astl}{\underset{\mathcal{L}_l}{\ast}}
\nc{\astj}{\underset{\mathcal{L}_j}{\ast}}
\nc{\astll}{\underset{\mathcal{L}_1}{\ast}}
\nc{\Yt}{Y^{\textrm{th}}}
\nc{\rmd}{{\rm d}}
\nc{\rmi}{{\rm i}}
\nc{\rme}{{\rm e}}

\DeclareMathOperator{\sh}{sh}
\DeclareMathOperator{\ch}{ch}
\DeclareMathOperator{\tnh}{th}
\DeclareMathOperator{\Tr}{Tr}

\nc{\End}{\mathrm{End}}

\numberwithin{equation}{section}

\begin{document}

\begin{center}{\Large \textbf{The spin Drude weight of the spin-1/2 $XXZ$ chain:\\
 An analytic finite size study
}}\end{center}

\begin{center}
Andreas Kl\"umper\textsuperscript{1} and
Kazumitsu Sakai\textsuperscript{2*}
\end{center}

\begin{center}
{\bf 1} Fakult\"at f\"ur Mathematik und Naturwissenschaften,\\
Bergische Universit\"at Wuppertal, 42097 Wuppertal, Germany
\\
{\bf 2} Department of Physics, Tokyo University of Science,\\
Kagurazaka 1-3, Shinjuku-ku, Tokyo 162-8601, Japan
\\[5mm]

* k.sakai@rs.tus.ac.jp
\end{center}

\begin{center}
April 25, 2019
\end{center}


\section*{Abstract}
{\bf
The Drude weight for the spin transport of the spin-1/2 $XXZ$ Heisenberg chain
in the critical regime is evaluated exactly for finite temperatures. We
combine the thermodynamic Bethe ansatz with the functional relations of type
$Y$-system satisfied by the row-to-row transfer matrices. This makes it
possible to evaluate the asymptotic behavior of the finite temperature spin
Drude weight with respect to the system size.  As a result, the Drude weight
converges to the results obtained by Zotos (Phys. Rev. Lett.
\textbf{82}, 1764 (1999)), however with very slow convergence upon increase of the system
size. This strong size dependence may explain that extrapolations from various
numerical approaches yield conflicting results.
}

\vspace{10pt}
\noindent\rule{\textwidth}{1pt}
\tableofcontents\thispagestyle{fancy}
\noindent\rule{\textwidth}{1pt}
\vspace{10pt}

%
\section{Introduction}
The Heisenberg spin-1/2 $XXZ$ chain is one of the most fundamental and
extensively investigated models describing low-dimensional quantum magnetism.
The model is exactly solvable due to the existence of infinitely many
nontrivial local conserved charges. The static properties, such as the energy
spectrum, thermodynamic quantities and even correlation functions, can be
exactly calculated by various versions of the Bethe ansatz (BA).  In contrast, the
evaluation of quantities related to dynamical properties is in general much
more difficult.

Nevertheless, by utilizing the integrability, several transport coefficients
of the model have been calculated within linear response theory. For instance,
for the thermal transport, the energy current itself is conserved
\cite{ZNP97}, and hence the thermal Drude weight, which is the weight of a
delta function contribution to the thermal conductivity, can be evaluated
\cite{KS02,KS03,Zotos17} by a generalized Gibbs ensemble.

The spin transport is a more intricate topic as the spin current -- in
contrast to the thermal current -- is not a conserved quantity. The spin Drude
weight at finite temperatures is the main topic of this paper.  For certain
anisotropy values of the $XXZ$ chain the finite temperature spin Drude weight
was evaluated using exact means by Zotos \cite{Zotos} 20 years
ago. These calculations consist of two steps. First, the spin Drude weight is
expressed in terms of the thermal expectation value of the curvature of the
energy eigenvalues with respect to the magnetic flux through the system. This
is a finite-temperature generalization of the Kohn formula
\cite{Kohn}. Second, by taking into account as carefully as possible the
behavior of string type solutions to the Bethe ansatz equations with respect
to the flux, the spin Drude weight is described by a set of non-linear
integral equations that may be viewed as extended thermodynamic Bethe ansatz
(TBA) equations \cite{Zotos}.  Originally this procedure has been developed
for the evaluation of the Drude weight for the Hubbard chain \cite{FK}. It is
based on the so-called string hypothesis according to which bound states are
described by equidistantly distributed Bethe rapidities with corrections that
are mostly but not always exponentially small for large system sizes.

The results of \cite{Zotos} indicate that the Drude weight of the $XXZ$ chain
in the critical regime is finite and monotonously decreases with increase of
temperature as well as with increase of the anisotropy parameter. It is
exactly zero at any finite temperature for the isotropic point and for the
massive regime.

There exist, however, several controversies about the spin Drude weight for
the $XXZ$ chain: various analytical or numerical studies yield 
contradictory results \cite{NMA,AG,MHCB,BFKS,HS,SPA09,SPA11,HPZ,CPC,SV}.  In
recent years, novel approaches \cite{Pro11,PI, Pro14,IMP,IN,PPS,UOKS} have
been developed based on the Mazur inequality, new conserved quasi-local
charges, and their charge-charge and current-charge correlation
functions. These can be calculated exactly by the Bethe ansatz or, at high
temperatures, by more elementary means. Specifically at high temperatures,
an optimal lower bound has been evaluated \cite{PI} which agrees with the
high-temperature asymptotics of Zotos' results \cite{Zotos,BFKS}. For general
temperature the extended TBA equations of \cite{Zotos} can be derived on the
basis of an optimal lower bound \cite{UOKS} complementing the derivation by
the generalized hydrodynamics developed recently in
\cite{BCNF,ADY,BVKM,DS,NBD}.  Interestingly, the spin Drude weight at any
finite temperatures exhibits a fractal dependence on the anisotropy parameter
of the model \cite{Pro11,PI,UOKS}.

Motivated by these results, we revisit this long-standing problem for the spin
Drude weight of the $XXZ$ chain.  We develop a completely different method for
evaluating Kohn's formula. We avoid the treatment of string corrections by use
of the $T$- and $Y$-systems \cite{KSS} which are the functional relations
satisfied by transfer matrices ($T$-functions) and certain combinations
thereof ($Y$-functions). All excitations are described by real excitation
parameters, i.e.~the zeros of the $T$-functions.  The curvature of the energy
levels can be expressed by the $Y$-functions, and their particular zeros
characterizing the state.  By insertion of the distribution of zeros, which
describes the thermal equilibrium, the thermal expectation value of the
curvature of the energy spectrum and hence the Drude weight is obtained. In
the thermodynamic limit, indeed the formulation of Zotos is obtained.
However, our approach allows to estimate the large-size asymptotics of the
spin Drude weight.  We find that the Drude weight strongly depends on the
system size, and very slowly converges to the result in the thermodynamic
limit.

The layout of this paper is as follows. In the next section, we briefly
formulate the spin transport within linear response theory. Also a
finite-temperature generalization of the Kohn formula is presented. In section
3, we give the $T$- and $Y$-system defined for the six-vertex model which is
the classical counterpart of the $XXZ$ chain.  In section 4, we show that the
energy for any excited state is expressed as the solution to the non-linear
integral equations (NLIEs) satisfied by the $Y$-functions. The
finite-temperature extension is described in section 5.  The analysis of the
NLIEs and the temperature and size-dependences of the Drude weight are
presented in section 5. The last section is devoted to a summary and
discussions. Some technical details are deferred to several appendices.

\section{Drude weight}

Let us consider the spin transport property of the spin-1/2 $XXZ$ chain on the
periodic lattice with sites labeled by $k=1,2,\dots, L$:
\begin{equation}
H=J\sum_{k=1}^L 
 \left(
  \sigma_{k+1}^+\sigma_{k}^-+\sigma_{k}^+\sigma_{k+1}^-
 +\frac{\Delta}{2}\sigma_k^z\sigma_{k+1}^z
\right).
\label{xxz}
\end{equation}
Here $\sigma_k^x$, $\sigma_k^y$, $\sigma_k^z$ are the Pauli matrices acting on
the $k$th site and $\sigma_k^{\pm}:=(\sigma_k^x\pm \rmi \sigma_k^y)/2$.  For
later convenience, we parameterize the anisotropy parameter $\Delta$ as
\begin{equation}
\Delta=\cos\gamma \quad (0\le\gamma\le \frac{\pi}{2})
\label{parameter}
\end{equation}
and set
\begin{equation}
\nu=\frac{\pi}{\gamma}  \quad (\nu\ge 2).
\label{nu}
\end{equation}
To consider the spin conductivity $\sigma_{\rm s}$ within linear response
theory, we utilize the Kubo formula \cite{Kubo,Mahan}
\begin{equation}
\sigma_{\rm s}:=\lim_{\omega\to 0}\Re \sigma_{\rm s}(\omega),\quad
\sigma_{\rm s}(\omega):=\lim_{\varepsilon\to +0}\frac{1}{L}
       \int_0^\infty \rmd t\, e^{-\rmi(\omega-\rmi\varepsilon)t}
           \int_0^{\beta}\rmd\lambda\, 
             \bra \mathcal{J}_{\rm s}(-t-\rmi\lambda)\mathcal{J}_{\rm s}\ket,
\label{cor}
\end{equation}
where $\bra \cdots \ket$ denotes the thermal expectation value, $\beta$ is the
reciprocal of temperature, $\beta=1/T$, and $\mathcal{J}_{\rm s}$ is the total
spin current given by
\begin{equation}
\mathcal{J}_{\rm s}=\sum_{k=1}^L j_k^{\rm s}=
-\rmi J \sum_{k=1}^L (\sigma_{k+1}^+\sigma_{k}^--\sigma_{k}^+\sigma_{k+1}^-),
\label{sc}
\end{equation}
which is naturally derived from the continuity equation:
\begin{equation}
\dot{S}_k^z=-(j_{k+1}^{\rm s}-j_{k}^{\rm s}),
\quad
\dot{S}_k^z=\frac{1}{2}\dot{\sigma}_k^z=\rmi\left[H,\frac{1}{2}\sigma_k^z\right].
\end{equation}

After some simple manipulation, the spin conductivity can be expressed as
\begin{equation}
\Re \sigma_{\rm s}(\omega)=\pi D \delta(\omega)+\sigma_{\rm s}^{\rm reg}(\omega).
\end{equation}
The singular part $\pi D \delta(\omega)$ denotes the ballistic
(dissipationless) contribution, whereas $\sigma_{\rm s}^{\rm reg}(\omega)$
describes the normal (dissipative) transport. The weight $D$ of the delta
function is called the spin Drude weight and is a characteristic of ballistic
transport.
In the thermodynamic limit $L\to\infty$, \eqref{cor} is reduced to
\begin{align}
D:=-\frac{1}{L}\left[\bra K \ket+2\sum_{\substack{m,n \\ E_m\ne E_n}}p_n 
\frac{|\bra m|\mathcal{J}_{\rm s}|n \ket|^2}{E_m-E_n}\right],
\label{drude}
\end{align}
where $p_n:=\rme^{-\beta E_n}/(\sum_m \rme^{-\beta E_m})$ is the Boltzmann
weight for the energy eigenvalue $E_n$ corresponding to the eigenstate $|n
\ket$, and $\bra K \ket$ is the thermal expectation value of the
kinetic energy term $K:=J\sum_k
(\sigma_{k+1}^+\sigma_{k}^-+\sigma_{k}^+\sigma_{k+1}^-)$.

In general, the direct evaluation of \eqref{drude} is a formidable task
especially for $L\gg 1$.  Fortunately, this expression can be transformed
into a more accessible form. Let us introduce a flux $\phi$ into the system
\eqref{xxz}, i.e.
\begin{equation}
H(\phi)=J\sum_{k=1}^L 
 \left(
  \rme^{\rmi\phi/L} \sigma_{k+1}^+\sigma_{k}^-
 +\rme^{-\rmi\phi/L}\sigma_{k}^+\sigma_{k+1}^-
 +\frac{\Delta}{2}\sigma_k^z\sigma_{k+1}^z
\right),
\label{xxz-phi}
\end{equation}
and perturbatively expand the energy $E_n(\phi)$ for \eqref{xxz-phi} in terms
of $\phi/L$:
\begin{equation}
E_n(\phi)=E_n-\frac{\phi}{L}\bra n |\mathcal{J}_{\rm s}|n \ket
-\left(\frac{\phi}{L}\right)^2\sum_{\substack{m,n \\ E_m\ne E_n}}
\frac{|\bra m|\mathcal{J}_{\rm s}|n \ket|^2}{E_m-E_n}
-\frac{1}{2}\left(\frac{\phi}{L}\right)^2\bra n | K | n\ket+o\left(\phi^2/L^2\right).
\end{equation}
Thus one finds that the Drude weight $D$ is identical to the thermal average
of energy level curvatures with respect to $\phi$:
\begin{equation}
D=\left. \lim_{L\to\infty}L \sum_n p_n 
\frac{\rmd^2 E_n(\phi)}{\rmd \phi^2}\right|_{\phi=0}.
\label{drude2}
\end{equation}
This formula is the finite-temperature generalization of Kohn's formula
\cite{Kohn}.

\section{$T$- and $Y$-systems}
%
As shown in \eqref{drude2}, to evaluate the finite-temperature Drude weight,
we must carefully take into account the thermal expectation value of the
second derivative of the energy with respect to the flux. To achieve this
systematically, we translate the original problem for the (1+1)-dimensional
quantum system into a problem for the two-dimensional classical lattice
system.  The lattice system corresponding to the $XXZ$ chain is the six-vertex
model whose weights are given by the following six non-zero elements of the
$R$-matrix $R(v)\in \End(V\otimes V)$ ($V$ denotes the two-dimensional vector
space $\mathbb{C}^2$ spanned by the spin-up state $|0\ket=\binom{1}{0}$ and
the spin-down state $|1\ket=\binom{0}{1}$):
\begin{equation}
R_{11}^{11}(v)=R_{22}^{22}(v)=\frac{[v+2]}{[2]},
\quad
R_{12}^{12}(v)=\rme^{\rmi\frac{\phi}{L}}\frac{[v]}{[2]},
\quad
R_{21}^{21}(v)=\rme^{-\rmi\frac{\phi}{L}}\frac{[v]}{[2]},
\quad
R_{12}^{21}(v)=R_{21}^{12}(v)=1,
\label{6-vertex}
\end{equation}
where $[v]:=\sin(\gamma v/2)/\sin\gamma$. The indices
can be interpreted as
\begin{equation}
R(v)|\alpha\ket\otimes|\beta\ket
=\sum_{\gamma,\delta}R^{\gamma\delta}_{\alpha \beta} (v) |\gamma\ket\otimes|\delta\ket.
\end{equation}
The $R$-matrix satisfies the Yang-Baxter equation (YBE)
\begin{equation}
R_{23}(v)R_{13}(u)R_{12}(u-v)=
R_{12}(u-v)R_{13}(u)R_{23}(v),
\label{ybe}
\end{equation}
where $R_{jk}(v)$ acts on $V_j\otimes V_k$. Note that $V_j$ means the copy of
$\mathbb{C}^2$ spanned by the $j$th state $|0\ket_j$ and $|1\ket_j$. The YBE
guarantees that the family of row-to-row transfer matrices
$T_1(v)\in \End(V^{\otimes L})$ constructed by
\begin{equation}
T_1(v)=\Tr_{V_0}
\left[
R_{0L}(i(v+\rmi))R_{0L-1}(\rmi(v+\rmi))\cdots R_{01}(\rmi(v+\rmi))
\right]
\label{transfer}
\end{equation}
consists of mutually commuting matrices for arbitrary spectral parameters $u$
and $v$:
\begin{equation}
[T_1(u),T_1(v)]=0.
\end{equation}
The original quantum system \eqref{xxz-phi} can be expressed as the
logarithmic derivative of $T_1(v)$ with respect to $v$:
\begin{equation}
H(\phi)= \left.\frac{A}{2\pi \rmi}
\frac{\partial}{\partial v}\log T_1(v) \right|_{v=-\rmi}-\frac{JL}{2}\Delta,
\quad A:=\frac{4\pi J\sin\gamma}{\gamma}.
\label{baxter}
\end{equation}
The transfer matrices $T_1(v)$ can be diagonalized by standard Bethe ansatz
techniques.  The resultant eigenvalues (simply denoted by $T_1(v)$) explicitly
read
\begin{align}
&T_1(v)=\varphi(v-\rmi)\frac{q(v+2\rmi)}{q(v)}\rme^{\rmi \frac{\phi}{2}}+
\varphi(v+\rmi)\frac{q(v-2\rmi)}{q(v)}\rme^{-\rmi\frac{\phi}{2}}, \nn \\
&\varphi(v):=\left(\frac{\sh\frac{\gamma}{2}v}{\sin\gamma}\right)^L,
\qquad q(v):=\prod_{k=1}^{m}\sh\frac{\gamma}{2}(v-v_k).
\label{dvf}
\end{align}
Here we have dropped an overall phase factor $\rmi^L \rme^{\rmi
  \phi(m/L-1/2)}$ in the expression of $T_1(v)$.  The
$m\in\{0,1,2,\dots,L/2\}$ unknown numbers $\{v_k\}$ in \eqref{dvf} are to be
determined from the Bethe ansatz equation (BAE):
\begin{equation}
\frac{\varphi(v_k+\rmi)}{\varphi(v_k-\rmi)}=-\rme^{\rmi\phi}
\frac{q(v_k+2\rmi)}{q(v_k-2\rmi)}.
\label{bae}
\end{equation}
Instead of solving the BAE \eqref{bae} directly, we introduce for convenience
a more general family of transfer matrices ($T$-functions) including the above
$T_1(v)$.
Let us consider the $T$-functions of the form \cite{KSS}
\begin{equation}
T_{n-1}(v):=\sum_{j=1}^{n}\rme^{\frac{\rmi\phi}{2}(n-2j+1)}
\varphi(v+\rmi(2j-n-1))\frac{q(v+\rmi n)q(v-\rmi n)}
{q(v+\rmi(2j-n-2))q(v+\rmi(2j-n))}.
\label{T-func}
\end{equation}
The following relations ($T$-system) are directly proven for any
$v\in\mathbb{C}$ and integers $1\le j\le n$,
\begin{equation}
T_{n-1}(v+\rmi j)T_{n-1}(v-\rmi j)=
T_{n+j-1}(v)T_{n-j-1}(v)+T_{j-1}(v+\rmi n)T_{j-1}(v-\rmi n),
\label{tsys1}
\end{equation}
where we set $T_{-1}(v)=0$ and $T_0(v)=\varphi(v)$.

{}From now on, for simplicity, we restrict ourselves to $\nu\in\mathbb{Z}_{\ge
  2}$ in \eqref{nu}.
In this case, the following relation holds
\begin{equation}
T_{\nu}(v)=2(-1)^m \cos\left(\frac{\nu \phi}{2}\right)
T_0(v+\rmi \nu)+T_{\nu-2}(v).
\label{tsys2}
\end{equation}
Combining the $T$-functions, we construct the $Y$-functions \cite{KSS}:
\begin{align}
&Y_j(v)=\frac{T_{j-1}(v)T_{j+1}(v)}{T_0(v+\rmi(j+1))T_0(v-\rmi(j+1))}
\qquad \text{for $1\le j \le \nu-2$}, \nn \db
&Y_{\nu-1}(v)=(-1)^m \rme^{-\rmi \frac{\nu \phi}{2}}
               \frac{T_{\nu-2}(v)}{T_0(v+\rmi \nu)}, \nn \db
&Y_{\nu}(v)=(-1)^m \rme^{\rmi \frac{\nu \phi}{2}}
               \frac{T_{\nu-2}(v)}{T_0(v+\rmi \nu)}.
\label{yfunc1}
\end{align}
Here we set $Y_0(v)=0$.
Using the $T$-system \eqref{tsys1} and \eqref{tsys2}, one can easily find that
the above relations are equivalent to the following
\begin{align}
&1+Y_j(v)=\frac{T_j(v+\rmi)T_j(v-\rmi)}{T_0(v+\rmi(j+1))T_0(v-\rmi(j+1))} 
\qquad \text{for $1\le j\le \nu-2$}, \nn \db
&(1+Y_{\nu-1}(v))(1+Y_{\nu}(v))=\frac{T_{\nu-1}(v+\rmi)T_{\nu-1}(v-\rmi)}
                                    {T_0(v+\rmi \nu)T_0(v-\rmi \nu)}.
\label{yfunc2}
\end{align}
Moreover we notice that the $Y$-functions \eqref{yfunc1} satisfy the
functional relations ($Y$-system),
\begin{align}
&Y_j(v+\rmi)Y_j(v-\rmi)=(1+Y_{j-1}(v))(1+Y_{j+1}(v))   
\quad \text{for $1\le j\le \nu-3$}, \nn \db
&Y_{\nu-2}(v+\rmi)Y_{\nu-2}(v-\rmi)=
(1+Y_{\nu-3}(v))(1+Y_{\nu-1}(v))(1+Y_{\nu}(v)) \quad 
\text{for $\nu\ge 3$},  
 \nn \db
&Y_{\nu-1}(v+\rmi)Y_{\nu-1}(v-\rmi)=
\rme^{-\rmi \nu\phi}(1+Y_{\nu-2}(v)),   \nn \db
&Y_{\nu}(v+\rmi)Y_{\nu}(v-\rmi)=
\rme^{\rmi \nu\phi}(1+Y_{\nu-2}(v)). 
\label{ysys}
\end{align}

In the next section, we transform the above functional relations \eqref{ysys}
into the non-linear integral equations (NLIEs) which determine the energy
eigenvalues of the $XXZ$ model \eqref{xxz-phi}.

%
\section{NLIEs for arbitrary excitations}

\begin{figure}[t]
\centering
\includegraphics[width=0.85\textwidth]{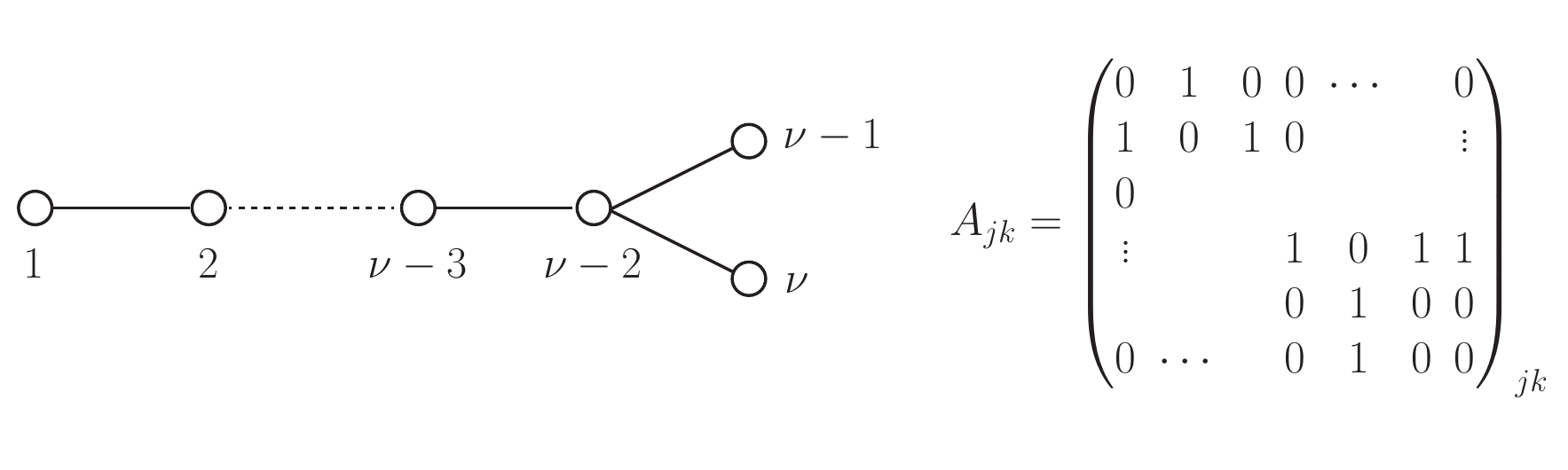}
\caption{The adjacency matrix $A_{jk}$ for type $D_{\nu}$. }
\label{a-matrix}
\end{figure}
From now on we consider the case that the system size is $L\in 4\mathbb{Z}$
and the number of BAE roots is $m=L/2=2\mathbb{Z}$.

The transfer matrix $T_1(v)$ \eqref{transfer} includes the Hamiltonian of the
$XXZ$ model \eqref{xxz-phi} via \eqref{baxter}.  The information on the energy
spectrum is embedded in analytical properties of the $T$- and
$Y$-functions. For instance, the ground state is characterized by those
$T$-functions for which none of the $T_j(v)$ ($j=1,\dots,\nu-1$) has any zero
in the strip $\Im v\in[-1,1]$ (which we call the physical strip).
On the other hand, any excited state is characterized by additional zeros
$\zeta_j^k$ satisfying the conditions $\Im \zeta_j^k\in [-1,1]$ and
\begin{equation}
Y_j(\zeta_j^k+\rmi)=
-1 \quad ( 1 \le k \le m_j \in \mathbb{Z}_{\ge 0};\,\, 1\le j\le \nu),
\label{sub1}
\end{equation}
or equivalently, from \eqref{yfunc2}
\begin{equation}
T_j(\zeta_j^k)=0 \quad (1\le j\le \nu-1), \quad  T_{\nu-1}(\zeta_{\nu}^k)=0.
\end{equation}
We give several remarks about the zeros $\{\zeta_j^k\}$. First, according to
usual conventions we may call the zeros $\{\zeta_j^k\}$ ($1\le j\le
\nu-1$) ``holes'' as they correspond to solutions of BA like equations, but
differ from BA rapidities. Second, by some numerical analysis and the form of
$T_{\nu-1}(v)$ in \eqref{T-func}, we find that $\{\zeta_{\nu}^k\}$ coincide
with the real parts of the odd-strings which are BA rapidities whose imaginary
parts are $\nu \rmi$. From the second equation in \eqref{yfunc2} we see
  that both $\{\zeta_{\nu-1}^k\}$ ($1\le k \le m_{\nu-1}$) and
$\{\zeta_{\nu}^k\}$ ($1\le k \le m_{\nu}=m_{\nu-1}$) are zeros of
$T_{\nu-1}(v)$.  For $\phi=0$, these zeros are degenerate: $\{\zeta_{\nu-1}^k
\}=\{\zeta_{\nu}^k\}$.  Finally, from \eqref{sub1} we obtain
\begin{equation}
-\log Y_j(\zeta^k_j+\rmi)=2\pi \rmi I_j^k \quad (I_j^k\in\mathbb{Z}+1/2).
\label{sub2}
\end{equation}
For this reason, the functions $-\log Y_j(v+\rmi)/(2\pi \rmi)$ are interpreted as
counting functions.
%

To find the solutions of $\eqref{ysys}$, we adopt the following procedure (see
appendix A for a detailed derivation).  First we take the logarithmic
derivative of \eqref{ysys} with respect to $v$. Second, performing the Fourier
transform and using Cauchy's theorem, we shift the contour of the integrals
on the left hand side to the real axis. Finally performing the inverse Fourier transform
and then integrating over $v$, we obtain the following NLIEs:
\begin{align}
&\log Y_j(v)=p_j(v)+g_j(v)+\sum_{l=1}^\nu K_{jl}\ast \log(1+Y_l)[v]
\quad
(1\le j \le \nu),
\label{Y-sys}
\end{align}
where 
\begin{equation}
K_{jk}(v):=A_{jk} s(v), \quad s(v):=\frac{1}{4\ch\frac{\pi}{2}v},
\end{equation}
and $A_{jk}$ denotes the adjacency matrix associated with the Dynkin diagram
of type $D_\nu$ (see Fig.~\ref{a-matrix}).  The symbol $*$ in \eqref{Y-sys}
denotes the convolution defined by $s\ast
f(v):=\int_{-\infty}^{\infty}s(v-x)f(x) \rmd x$.  The leading terms $p_j(v)$
and $g_j(v)$ in \eqref{Y-sys} are, respectively, given by
\begin{align}
&p_j(v):=(\delta_{j1}+\delta_{j2}\delta_{\nu2})( L p(v)+ s'(v-\rmi)\psi)-
                      \delta_{j \nu-1} \frac{\nu}{2} \phi \rmi
+\delta_{j \nu} \frac{\nu}{2} \phi \rmi,\nn \\
&g_j(v):=\sum_{l=1}^\nu A_{j l} \sum_{k=1}^{m_l}p(v-\zeta_l^k),
\label{leading}
\end{align}
where
\begin{equation}
p(v):=\log\tnh \frac{\pi}4 v.
\end{equation}
Here the artificial parameter $\psi$, which should be set to zero after all
calculations, has been introduced in the leading term $p_j(v)$ for later
convenience.
The NLIEs \eqref{Y-sys} together with the subsidiary conditions \eqref{sub2}
give the solutions to the $Y$-system \eqref{ysys}.

Modifying the integration contours in the convolutions so that they surround
the
parameters $\zeta_l^k+\rmi$ $(1\le k \le m_l$) in clockwise manner (see
Fig.~\ref{contour}), we can reduce the NLIEs to simpler forms (see appendix A
in detail):
\begin{equation}
\log Y_j(v)=p_j(v)+\sum_{l=1}^{\nu}K_{jl}\astl \log(1+Y_l)[v],
\label{modified}
\end{equation}
where $\mathcal{L}_l$'s ($l=1,\dots,\nu$) denote the modified contours.  For
the ground state, $\mathcal{L}_l$'s are just straight lines.

\begin{figure}[t]
\centering
\includegraphics[width=0.55\textwidth]{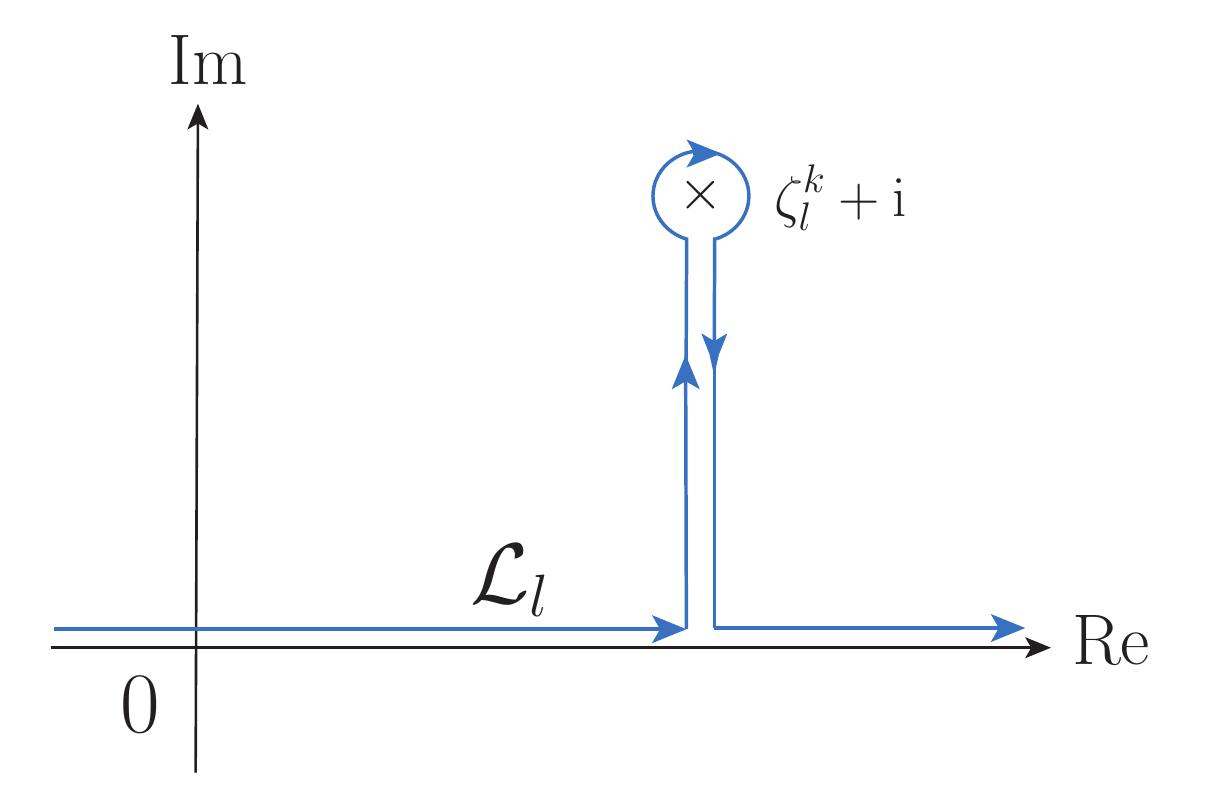}
\caption{Depiction of a modified contour for $1+Y_l(v)$ in
  \eqref{modified}. The contour is modified such that it encircles
  $\zeta_l^k+\rmi$ $(1\le k \le m_{l})$ in clockwise direction.}
\label{contour}
\end{figure}

Applying the same procedures to \eqref{yfunc2} (see also appendix A for the
derivation), and using the identity \eqref{baxter}, one obtains the energy
eigenvalues for arbitrary excited states:
\begin{align}
E(\phi)&= E_{\rm gs}+A \sum_{j=1}^{\delta_{\nu 2}+1}\left[\sum_{k=1}^{m_j}s(\zeta_j^k)+
 \frac{1}{2\pi \rmi} s'\ast\log(1+Y_j)[-\rmi]\right]  \nn \\
&=E_{\rm gs}+\frac{A}{2\pi \rmi}\sum_{j=1}^{\delta_{\nu 2}+1}s'
\astj\log(1+Y_j)[-\rmi],
\label{spectrum}
\end{align}
where $E_{\rm gs}$ is the ground state energy given by
\begin{equation}
E_{\rm gs}=\frac{JL}{2}\Delta-A L \int_{-\infty}^{\infty}
a(v)s(v)\rmd v, \quad a(v):=\frac{\sin\gamma}{2\nu(\ch \gamma v-\cos\gamma)}.
\label{gs}
\end{equation} 
All the information of the state is included in the set of the contours
$\{\mathcal{L}_l\}$: each set $\{\mathcal{L}_l\}$ defines a different state.

Thus the curvature of the energy spectrum (see \eqref{drude2}) is given by
\begin{equation}
\left.L \frac{d^2 E}{d\phi^2}\right|_{\phi=0}
=\frac{AL}{2\pi \rmi}\sum_{j=1}^{\delta_{\nu2}+1}  s'
\left.\astj\der_\phi^2\log(1+Y_j)[-\rmi]\right|_{\phi=0}.
\label{drude3}
\end{equation}

%
\section{Finite temperatures}
%
Each set of
parameters $\{\zeta_j^k\}$ gives a specific energy level via
\eqref{spectrum}.
For $L\gg 1$, the
distributions corresponding to equilibrium states are
characterized by the condition of minimizing the free energy. The
thermodynamic quantities such as the internal energy $\bra E \ket$ and $\bra L
\rmd^2 E/\rmd \phi^2 \ket$, which is the quantity we like to calculate, can be
evaluated by inserting the characteristic distributions into \eqref{spectrum}
and \eqref{drude3}.

Let us determine for $L\gg 1$ the
distributions describing the equilibrium.  The quantity
$1/L(\zeta_j^{k+1}-\zeta_j^k)$ yields the density of the parameters
  $\zeta_j^k$.  For all $j<\nu$ these densities will appear to have the
  meaning of hole densities of strings in the TBA approach, and for $j=\nu$
  the distribution function has the meaning of the density function of
  odd-strings.  We simply denote all such densities by $\rho_{j}^{\rm
    h}(v)$. For all $j$ these functions describe the distribution of $\zeta_j^k$
\begin{equation}
\rho_j^{\rm h}(\zeta_j^k)
=\lim_{L\to\infty}\frac{1}{L(\zeta_j^{k+1}-\zeta_j^k)}\,\,\,(1\le j\le\nu), 
\label{def-hole1}
\end{equation}
On the other hand the quantity
$\lim_{L\to\infty}(I_j^{k+1}-I_j^k)/(L(\zeta_j^{k+1}-\zeta_j^k))$ gives the
total density of holes and ``particles'':
\begin{equation}
\lim_{L\to\infty}
\frac{I_j^{k+1}-I_j^k}{L(\zeta_j^{k+1}-\zeta_j^k)}=
\rho_j^{\rm h}(\zeta_j^k)+\rho_j (\zeta_j^k),
\label{def-hole2}
\end{equation}
where $I_j^k$ denote the quantum numbers occurring in \eqref{sub2}.
The substitution of \eqref{sub2} and \eqref{Y-sys} into \eqref{def-hole2}
yields
\begin{align}
\rho_{j}^{\rm h}(v)+\rho_j(v)=-\lim_{L\to\infty}
\frac{1}{2\pi \rmi L}\der_v\log Y_j(v+\rmi)  
=s_j(v)+\sum_{l=1}^{\nu} K_{jl}\ast\rho_l^{\rm h}[v],
\label{hole}
\end{align}
where 
\begin{equation}
s_j(v):=(\delta_{j1}+\delta_{j2}\delta_{\nu2})s(v),
\label{driving}
\end{equation}
and we have set the parameter $\psi=0$ in \eqref{leading}.  By use of the
density functions $\rho_j^{\rm h}(v)$ \eqref{def-hole1} 
and the energy spectrum \eqref{spectrum}, 
one finds that the energy eigenvalues per site $e$ can be written as
\begin{align}
&e=e_{\rm gs}+
A\int_{-\infty}^{\infty}s(v)\left\{\rho_1^{\rm h}(v)+
\delta_{\nu 2}\rho^{\rm h}_{\nu}(v)\right\} \rmd v,
\label{eps}
\end{align}
where  $e_{\rm gs}:= \lim_{L\to\infty} E_{\rm gs}/L$ is the ground state energy
per site (see \eqref{gs}).
The entropy per site $s$ is evaluated by taking the logarithm of the number
of ways to take $\rho_j^{\rm h}(v)dv$ objects from a sequence of $(\rho_j^{\rm
  h}(v)+\rho_j(v))dv$ objects:
\begin{equation}
s=\sum_{j=1}^\nu \int_{-\infty}^\infty
\left[
(\rho_j(v)+\rho_j^{\rm h}(v))\log(\rho_j(v)+\rho_j^{\rm h}(v)) dv
-\rho_j(v)\log \rho_j(v)-\rho_j^{\rm h}(v)\log\rho_j^{\rm h}(v)
\right]\rmd v.
\end{equation}
Minimizing the free energy, $\delta f=\delta(e-T s)=0$, we derive the so-called
TBA equations 
determining the density functions
$\rho_j^{\rm h}(v)$ ($1\le j\le\nu$) 
which characterize the equilibrium state:
\begin{equation}
\log \eta_j(v)=-\beta A s_j(v)+\sum_{l=1}^{\nu} K_{j l}\ast\log(1+\eta_l)[v],
\label{TBA}
\end{equation}
where we have defined
\begin{equation}
\eta_j(v):=\frac{\rho_j^{\rm h}(v)}{\rho_j(v)} \quad (1\le j\le\nu).
\end{equation}

This equation exactly agrees with the one derived by the string hypothesis
\cite{TS,Takbook}. However, we emphasize that our formula does not rely 
on the string hypothesis, but only on the simple analytical assumption 
explained previously. In particular, the parameters $\zeta_j^k$ are real
and hence no finite size dependent corrections of imaginary parts appear.
The comparison of \eqref{TBA} with \eqref{hole} leads to
\begin{equation}
\rho_j^{\rm h}(v)=-\frac{1}{A}\der_\beta \log(1+\eta_j(v)) \quad (1\le j\le \nu).
\label{rhoh}
\end{equation}
Substituting these density functions into the NLIEs \eqref{Y-sys}, we obtain
the ``thermal" NLIEs (not to be confused with the TBA equations):
\begin{align}
\log \Yt_j(v)&=p_j(v)+g^{\rm th}_j(v)+\sum_{l=1}^\nu K_{jl}
\ast \log(1+\Yt_l)[v] \nn \\
&=p_j(v)+\sum_{l=1}^\nu K_{jl}
\astl \log(1+\Yt_l)[v] ,
\label{thermal}
\end{align}
with
\begin{equation}
g^{\rm th}_j(v)=L \sum_{l=1}^{\nu} A_{jl}\, p\ast\rho_l^{\rm h}[v].
\label{p-func}
\end{equation}
Thus combining \eqref{thermal}, \eqref{drude3} and \eqref{drude2}, the Drude
weight for finite temperature is given by
\begin{equation}
D=\lim_{L\to\infty}\left\bra L\frac{\rmd^2 E(\phi)}{\rmd \phi^2} 
\right\ket=\lim_{L\to\infty}\frac{AL}{2\pi \rmi} 
\sum_{j=1}^{1+\delta_{\nu2}}
\left.s'\astj\der_\phi^2\log(1+\Yt_j)[-\rmi]\right|_{\phi=0,\psi=0}.
\label{drude-finite}
\end{equation}

\section{Analysis of the Drude weight}
In this section, we derive a manageable formula for the Drude weight for
finite temperature by analyzing the thermal NLIEs \eqref{thermal} together
with the TBA equations \eqref{TBA}. We evaluate both temperature and size
dependences of the Drude weight.

\subsection{Analysis of the NLIEs}

By using the techniques of the dressed function formalism as shown in appendix
B, the Drude weight is rewritten as \eqref{dressed-drude}.  Let us simplify
this equation step by step.  First, from \eqref{thermal} we find that
\begin{equation}
\left.\der_\phi \log \Yt_j(v)\right|_{\phi=0,\psi=0}=
-\frac{\nu i}{2}\delta_{j\nu-1}+\frac{\nu i}{2}\delta_{j\nu}.
\end{equation}
Furthermore, the relation $\Yt_{\nu}(v)=\Yt_{\nu-1}(v)$ for $\phi=0$ reduces
\eqref{dressed-drude} to
\begin{equation}
D=\left.-\frac{AL\nu^2}{4\pi \rmi}\int_{\mathcal{L}_\nu} \rmd v
\frac{\der_{\psi} \log(1+\Yt_\nu(v))}{1+\Yt_\nu(v)}\right|_{\phi=0,\psi=0}.
\end{equation}

In the following, as already done above, we carry out derivatives with respect
to $\phi$ and/or $\psi$ and finally set these parameters to zero, which is
sometimes done implicitly by omitting the symbols $\phi=0$ and $\psi=0$ for
brevity.
Modifying the integrand as
\begin{equation}
\frac{\der_\psi \log (1+Y)}{1+Y}=
\frac{Y \der_\psi \log Y}{(1+Y)^2}
=-\frac{\der_\psi \log Y}{\der_v \log Y}\der_v \left(\frac{1}{1+Y}\right),
\end{equation}
and integrating by parts we find
\begin{equation}
D=\frac{AL\nu^2}{4\pi \rmi}
\left[\left.
\frac{\der_{\psi} \log \Yt_\nu(v)}{\der_v\log \Yt_\nu(v)}
\frac{1}{1+\Yt_\nu(v)} \right|_{-\infty}^\infty     
- \int_{\mathcal{L}_{\nu}} 
\frac{\rmd v}{1+\Yt_\nu(v)} 
\der_v
\left\{
\frac{\der_{\psi} \log \Yt_\nu(v)}{\der_v\log \Yt_\nu(v)}
\right\}
\right].
\end{equation}
The factor $1/(1+\Yt_\nu(v))$ in the integrand of the above equation has poles
at $v=\zeta_\nu^k+\rmi$ with residues
\begin{equation}
\frac{1}{{\Yt_\nu}'(\zeta_\nu^k+\rmi)}
=\frac{1}{\Yt_\nu(\zeta_\nu^k+\rmi)\log^{\prime}\Yt_\nu(\zeta_\nu^k+\rmi)}
=-\frac{1}{\log^{\prime}\Yt_\nu(\zeta_\nu^k+\rmi)}.
\end{equation} 
Using Cauchy's theorem, we modify the integration contour $\mathcal{L}_\nu$
to the straight line (see Fig.~\ref{contour}). Then performing again an
integration by parts, which cancels the surface terms, we arrive at
\begin{align}
D&=
-\frac{AL\nu^2}{2}\sum_{k=1}^{m_{\nu}}
\left.\frac{1}{\der_v\log \Yt_\nu(v)} 
\der_v\left\{
\frac{\der_{\psi} \log \Yt_\nu(v)}{\der_v\log \Yt_\nu(v)}
\right\}\right|_{v=\zeta_\nu^k+\rmi} \nn \\
&\qquad\qquad\qquad
+\frac{AL\nu^2}{4\pi \rmi}
\int_{-\infty}^\infty \rmd v
\der_v\left\{\frac{1}{1+\Yt_\nu(v)}\right\}
\frac{\der_{\psi} \log \Yt_\nu(v)}{\der_v\log\Yt_\nu(v)}\nn \\
&=-\frac{AL^2\nu^2}{2}\int_{-\infty}^{\infty}\rmd v
\frac{\rho^{\rm h}_\nu(v)}{\der_v\log \Yt_\nu(v+\rmi)} 
\der_v\left\{
\frac{\der_{\psi} \log \Yt_\nu(v+\rmi)}{\der_v\log \Yt_\nu(v+\rmi)}
\right\} \nn \\
&\qquad\qquad\qquad
+\frac{AL\nu^2}{4\pi \rmi}
\int_{-\infty}^\infty \rmd v
\der_v\left\{\frac{1}{1+\Yt_\nu(v)}\right\}
\frac{\der_{\psi} \log \Yt_\nu(v)}{\der_v\log\Yt_\nu(v)}.
\label{drude-finite2}
\end{align}
Note  that we have implicitly taken $\phi\to0$ and $\psi\to 0$
as explained above.

\begin{figure}[t]
\centering
\includegraphics[width=0.99\textwidth]{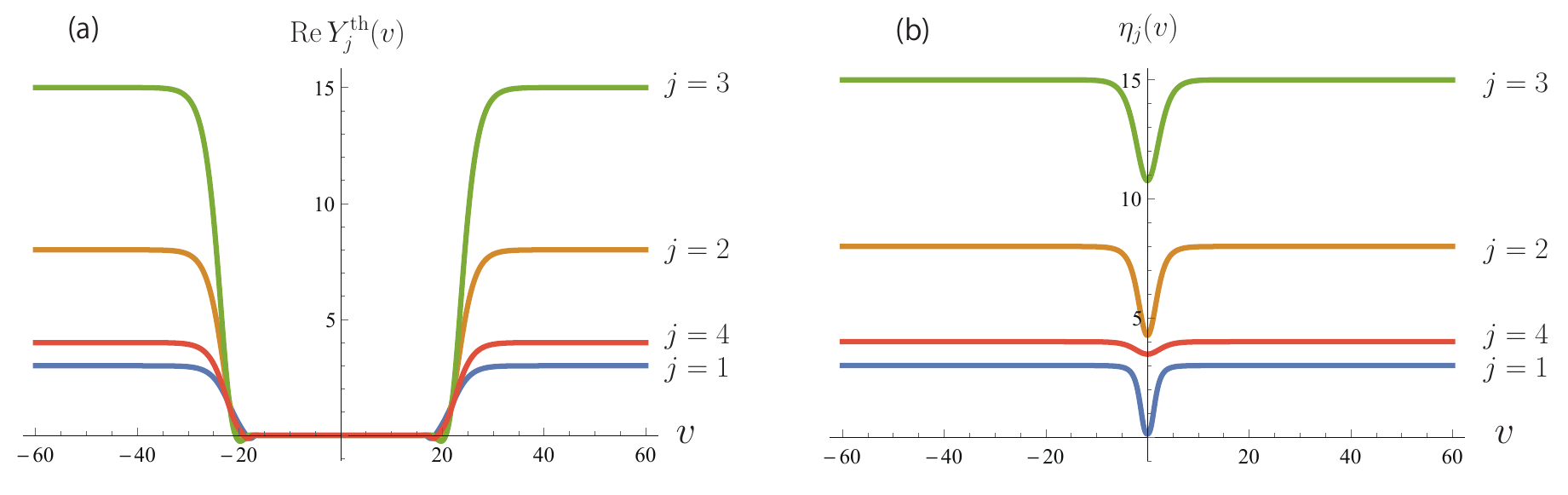}
\caption{The behavior of the functions $\Yt_j(v)$ and $\eta_j(v)$ satisfying
  \eqref{thermal} and \eqref{TBA} for $\nu=5$, $\beta J=1$ and $L=10^6$. The
  functions $\Yt_j(v)$ show a sharp crossover at about the points $v=\pm
  v_L=\pm (\nu/\pi)\log L\simeq 22$.  On the other hand, the function
  $\eta_j(v)$ smoothly vary in a small region.}
\label{y-function}
\end{figure}
Let us consider the thermodynamic limit $L\to\infty$. For $L\gg 1$, by some
simple analysis of the thermal NLIEs \eqref{thermal}, one observes that the
$Y$-functions $\Yt_j(v)$ show a sharp crossover at about the points
$v=\pm v_L:=\pm (\nu/\pi)\log L$:
\begin{equation}
\Yt_j(v)\simeq\begin{cases}
j(j+2) & (|v|>v_L) \nn \\
0 & (|v|<v_L)
\end{cases}
\  (j<\nu-1),
\ 
\Yt_{\nu-1}(v)=\Yt_{\nu}(v)=\begin{cases}
\nu-1 & (|v|>v_L) \nn \\
0 & (|v|<v_L)
\end{cases}.
\label{cross}
\end{equation}
See Fig.~\ref{y-function} (a) for the
behavior of $\Yt_j(v)$ for $\nu=5$, $\beta J=1$ and $L=10^6$.
On the other hand, for arbitrary finite temperatures the behavior of the
functions $\eta_j(v)$ is much smoother than that of $\Yt_j(v)$. See
Fig.~\ref{y-function} (b) for the behavior of $\eta_j(v)$ for $\nu=5$ and
$\beta J=1$.  Generally, $\eta_j(v)$ smoothly vary in a narrow region,
typically $|v|<v_\beta$ where $v_\beta\simeq (\nu/\pi) \log(1+\beta) \,\ll v_L$
and become constant outside this region.
\begin{equation}
\eta_j(v)\simeq\begin{cases}
j(j+2) & (j<\nu-1) \\
\nu-1  & (j=\nu-1,\, j=\nu)
\end{cases}
\quad (|v|>v_{\beta}).
\label{asymptotics}
\end{equation}
Thus to analyze \eqref{drude-finite2}, one must carefully take into account
this behavior of $\Yt_j(v)$ and $\eta_j(v)$.  First we consider $\der_v
\log\Yt_j(v)$.  Thanks to \eqref{TBA}, \eqref{rhoh} and \eqref{thermal}, we
identify
\begin{equation}
\der_v \log \Yt_j(v+\rmi)=
\frac{2\pi \rmi L}{A}\der_\beta \log \eta_j(v)+
\sum_{l=1}^\nu K_{jl}\ast\frac{\der_v \log \Yt_l}{1+1/\Yt_l}[v+\rmi].
\label{Yv}
\end{equation}
Due to the crossover behavior, the second term on the right hand side of the
above equation does not contribute for $|v|<v_L$, and therefore
\begin{equation}
\der_v \log \Yt_j(v+\rmi)\simeq
\frac{2\pi \rmi L}{A}\der_\beta \log \eta_j(v) \quad \text{for $|v|<v_L$}.
\label{Yv2}
\end{equation}
Though both functions $\der_v \log\Yt_j(v)$ and $(2\pi \rmi L/A)\der_\beta \log \eta_j(v)$
exponentially converge to zero as $v\to\pm \infty$, 
the decay rates are, in general, different due to the
$O(1)$ contribution of the second term in \eqref{Yv} for $|v|>v_L$.

Next we consider $\der_\psi \log \Yt_j(v)$. Using the same technique to
derive \eqref{drude-finite2}, we have
\begin{align}
\der_\psi \log \Yt_j(v+\rmi)
=&s'_j(v)-2\pi \rmi
L\sum_{l=1}^{\nu} \int_{-\infty}^{\infty}\rmd x
K_{jl}(v-x)\frac{\der_\psi \log \Yt_l(x+\rmi)}
                {\der_x\log \Yt_l(x+\rmi)}\rho_l^{\rm h}(x)\nn\\
&+\sum_{l=1}^\nu K_{jl}\ast\frac{\der_\psi \log\Yt_{\nu}}{1+1/\Yt_l}[v+\rmi].
\label{Yp0}
\end{align}
Inside the region $|v|<v_L$, the same argument to obtain \eqref{Yv} is also
applicable: we can ignore the third term on the right hand side of the above
equation. The insertion of \eqref{rhoh} and \eqref{Yv2} into the above leads
to
\begin{equation}
\der_\psi \log \Yt_j(v+\rmi)
=s_j'(v)+\sum_{l=1}^{\nu}\int_{-\infty}^{\infty}\rmd x\,K_{jl}(v-x)
\frac{\der_{\psi}\log \Yt_l(x+\rmi)}{1+1/\eta_l(x)}.
\end{equation}
Comparing this equation with the equation derived by taking the derivative of
\eqref{TBA} with respect to $v$, we find
\begin{equation}
\der_\psi \log  \Yt_j(v+\rmi)=-\frac{1}{\beta A} \der_v \log\eta_j(v)
\quad \text{for $|v|<v_L$}.
\label{Yp2}
\end{equation}
To describe $\der_{\psi}\log \Yt_j(v)$ in the whole region, we must consider
the correction term:
\begin{equation}
\der_\psi \log \Yt_j(v+\rmi)
=-\frac{1}{\beta A} \der_v \log\eta_j(v)+
\sum_{l=1}^\nu K_{jl}\ast\frac{\der_\psi \log\Yt_l}{1+1/\Yt_l}[v+\rmi].
\label{Yp}
\end{equation}
Again, both quantities $\der_{\psi}\log \Yt_j(v)$ and $-(1/\beta A) \der_v
\log\eta_j(v)$ exponentially converge to zero as $v\to\pm\infty$ with
different exponents. For the ratios, however, we find the relation
\begin{equation}
\frac{\der_\psi\log\Yt_j(v+\rmi)}{\der_v\log \Yt_j(v+\rmi)}\simeq
-\frac{1}{2\pi \rmi L\beta}\frac{\der_v\log\eta_j(v)}{\der_\beta\log \eta_j(v)}
\label{ratio}
\end{equation}
holds in the whole region, which follows from \eqref{Yv2} and \eqref{Yp2}
together with the fact that the driving terms of the linear integral equations
\eqref{Yv} and \eqref{Yp} converge exponentially to zero with the same
exponent for $|v|>v_\beta \,(\ll v_L)$.

Now we evaluate the Drude weight \eqref{drude-finite2} in the thermodynamic
limit.  As for the first term in the second equation of \eqref{drude-finite2}
  (let us denote it as $D_1$), the ratio of derivatives of $\log Y_\nu$ can be replaced
by those of $\log\eta_\nu$:
\begin{equation}
D_1=\frac{AL\nu^2}{4\pi \rmi\beta}\int_{-\infty}^{\infty}\rmd v
\frac{\rho^{\rm h}_\nu(v)}{\der_v\log \Yt_\nu(v+\rmi)} 
\der_v\left\{
\frac{\der_{v} \log \eta_\nu(v)}{\der_\beta\log \eta_\nu(v)}
\right\}.
\end{equation}
Since the second factor in the above integrand converges rapidly to zero for
$|v|> v_{\beta} \,(\ll v_L)$, one can neglect the behavior of the first factor
around $v=\pm v_L$. Consequently, we can replace the denominator of the first
factor by use of \eqref{Yv2}:
\begin{align}
D_1&=\frac{A}{8\gamma^2\beta}\int_{-\infty}^{\infty}\rmd v
\frac{\eta_\nu(v)}{1+ \eta_\nu(v)} 
\der_v\left\{
\frac{\der_{v} \log \eta_\nu(v)}{\der_\beta\log \eta_\nu(v)}
\right\} \nn \\
&=\frac{A}{4\gamma^2\beta}
\frac{\nu-1}{\nu} 
\frac{\der_{v} \log \eta_\nu(\infty)}{\der_\beta\log \eta_\nu(\infty)}
-\frac{A}{8\gamma^2\beta}\int_{-\infty}^{\infty}\rmd v
\frac{\eta_\nu(v)\left\{\der_{v} \log \eta_\nu(v)\right\}^2}
{\left\{1+ \eta_\nu(v)\right\}^2\der_\beta\log \eta_\nu(v)},
\label{D1}
\end{align}
where we have also substituted \eqref{rhoh} and the asymptotic value
\eqref{asymptotics}.
For the second term in \eqref{drude-finite2} (denoting it as $D_2$), we notice
that the first factor in the integrand is zero except for a small region
around $v=\pm v_L$. Hence the second factor can be replaced by a
ratio of derivatives of
$\log\eta$ functions. This ratio is in fact constant for $|v|>v_{\beta} \,(\ll
v_L)$, hence we obtain
\begin{equation}
D_2=\frac{-A}{4\gamma^2\beta}
\frac{\der_{v} \log \eta_\nu(\infty)}{\der_\beta\log \eta_\nu(\infty)}
\int_{0}^{\infty}\rmd v\,
\der_v\left(\frac{1}{1+1/\Yt_{\nu}(v)}\right)
=\frac{-A}{4\gamma^2\beta}
\frac{\nu-1}{\nu} 
\frac{\der_{v} \log \eta_\nu(\infty)}{\der_\beta\log \eta_\nu(\infty)},
\end{equation}
where we have inserted \eqref{cross}. In consequence, the surface term in
\eqref{D1} is exactly cancelled by the second term $D_2$, and then
\begin{equation}
D=-\frac{A}{8\gamma^2\beta}\int_{-\infty}^{\infty}\rmd v
\frac{\eta_\nu(v)\left\{\der_{v} \log \eta_\nu(v)\right\}^2}
{\left\{1+ \eta_\nu(v)\right\}^2\der_\beta\log \eta_\nu(v)}.
\label{zotos}
\end{equation}
This expression exactly coincides with the Drude weight derived by Zotos
\cite{Zotos}.  Especially for the $XY$ (free fermion) model ($\nu=2$), all the
functions $\Yt_j(v)$ and $\eta_j(v)$ are explicitly given by $\log
\Yt_j(v)=p_j(v)$ and $\log \eta_j(v)=-\beta s_j(v)$.
Then the Drude weight reads
\begin{equation}
D(\nu=2)=\frac{\beta J^2}{\pi}\int_{-\pi/2}^{\pi/2}
 \rmd p\,\frac{\sin^2 p}{\ch^2(\beta J\cos p)}.
\label{xy}
\end{equation}
This quantity is nothing but $\bra -K \ket/L$ as expected in
\eqref{drude}. (Note that the spin current $\mathcal{J}_{\rm s}$ \eqref{sc} is
a conserved quantity for the $XY$ case.)
%
\subsection{Numerical evaluation}
%
\begin{figure}[ttt]
\centering
\includegraphics[width=0.7\textwidth]{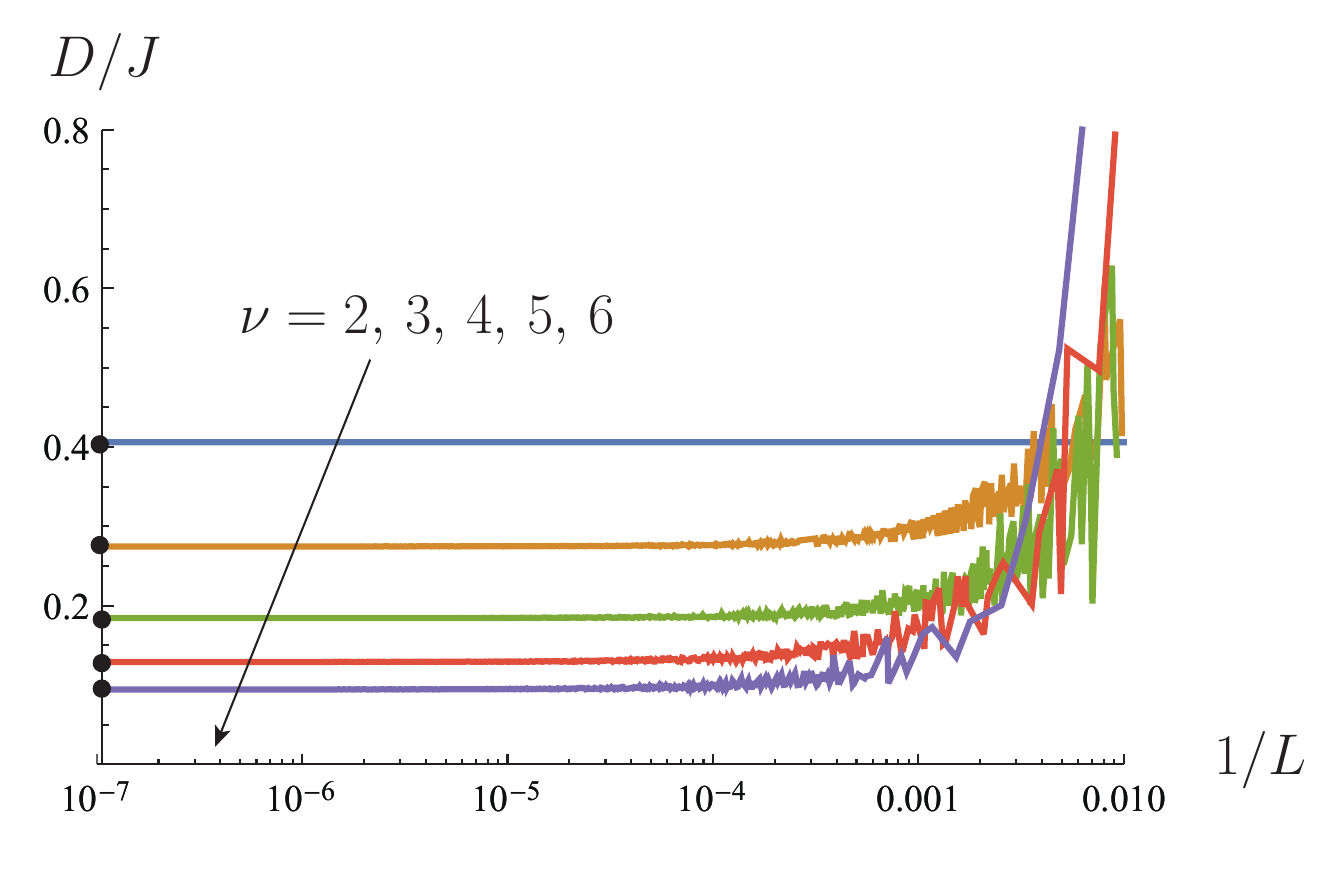}
\caption{The large-size asymptotics of the Drude weight $D$
  \eqref{drude-finite2} for $2\le \nu\le 6$ and fixed
    temperature $\beta J=1$. The Drude weight strongly depends on the system
  size even for large system size $L\simeq 10^3$, and quite slowly converges
  to the results derived by Zotos \cite{Zotos} (see also \eqref{zotos}) which
  are depicted by filled circles. In other words, the Drude weight sensitively
depends on the number of holes especially for small $L$.
In contrast, for the $XY$ chain ($\nu=2$),
  $D$ does not show such a strong size dependence.  The size dependence
  becomes more significant with the increase of
  $\Delta=\cos\pi/\nu$.
}
\label{drude-L}
\end{figure}
As shown in the previous subsection, the Drude weight $D$ for arbitrary system
size $L$ is given by \eqref{drude-finite2}, and converges  in
  the thermodynamic limit $L\to\infty$ to the result derived by Zotos
  \eqref{zotos}.  Here we evaluate \eqref{drude-finite2} numerically and
examine how this quantity converges to the results in the limit $L\to\infty$.
This can be achieved by numerically solving the NLIEs \eqref{thermal},
\eqref{Yv} and \eqref{Yp0} together with the TBA equations \eqref{TBA} and
\eqref{rhoh} which determine the hole densities $\rho_j^{\rm h}(v)$ for the
equilibrium state.
For finite size $L$ we use discrete distributions that approximate the
  continuous densities as closely as possible.
\begin{figure}[ttt]
\centering
\includegraphics[width=0.7\textwidth]{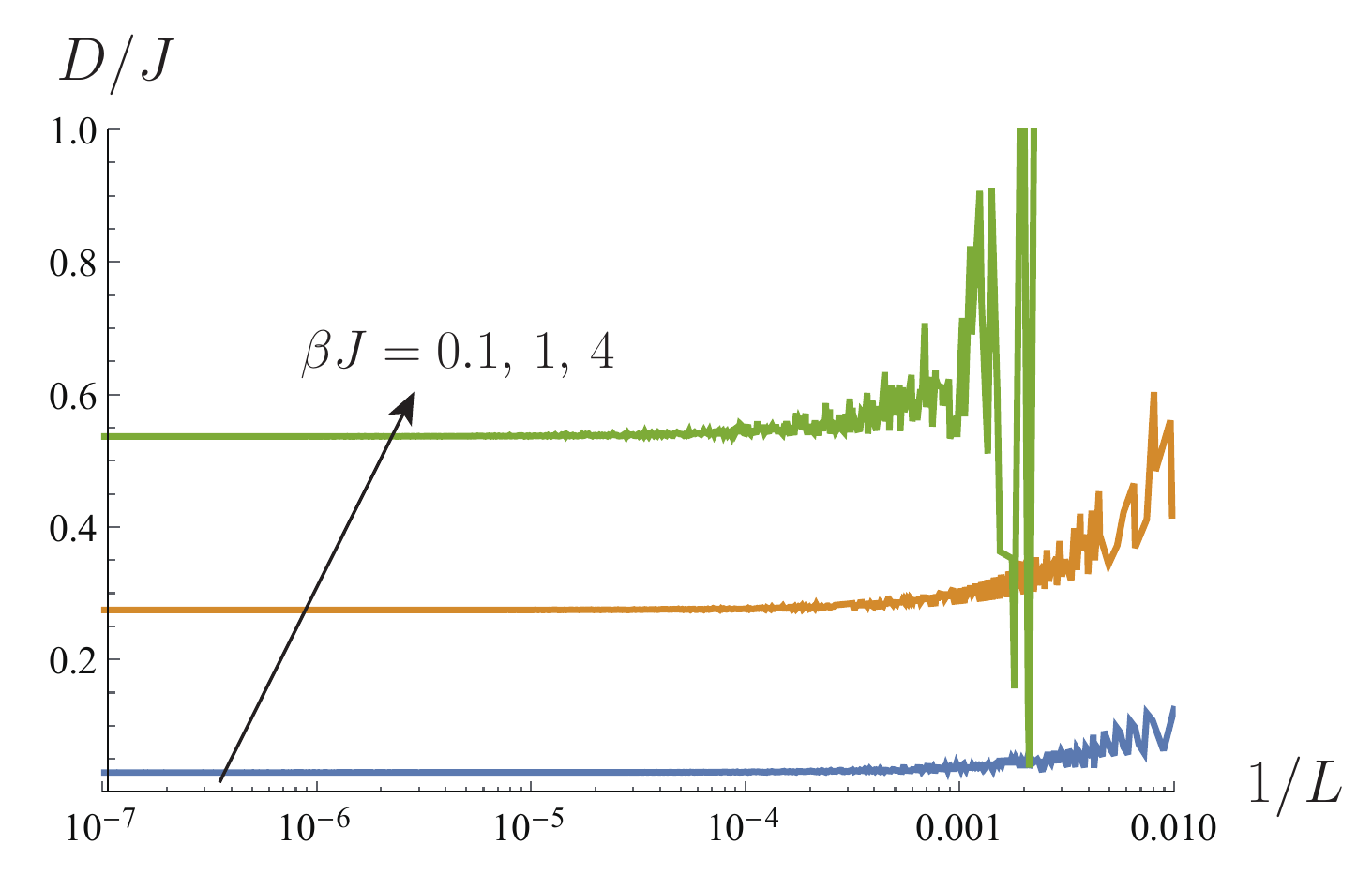}
\caption{The large size-asymptotics of the Drude weight $D$ for $\nu=3$ and
  different temperatures $\beta J=0.1,\, 1,\, 4$. With decrease
  of the temperature, the size dependence becomes more prominent.}
\label{drude-nu}
\end{figure}

In Fig.~\ref{drude-L}, the large-size asymptotic behavior of the Drude weight
$D$ is shown for various interaction strengths $\Delta=\cos\pi/\nu$ ($2\le
\nu \le 6$) with fixed temperature $\beta J=1$. We observe that $D$
sensitively depends on the system size (or equivalently on the number of
holes) even for $L\simeq 10^3$.  In contrast to this, for the $XY$ (free
fermion) model ($\nu=2$), where the Drude weight in the thermodynamic limit is
simply given by the bulk quantities \eqref{xy}, $D$ does not exhibit such a
strong size dependence.  In general, the bulk quantities such as the internal
energy quickly converge to the results for the thermodynamic limit
\eqref{eps}.

\begin{figure}[ttt]
\centering
\includegraphics[width=0.7\textwidth]{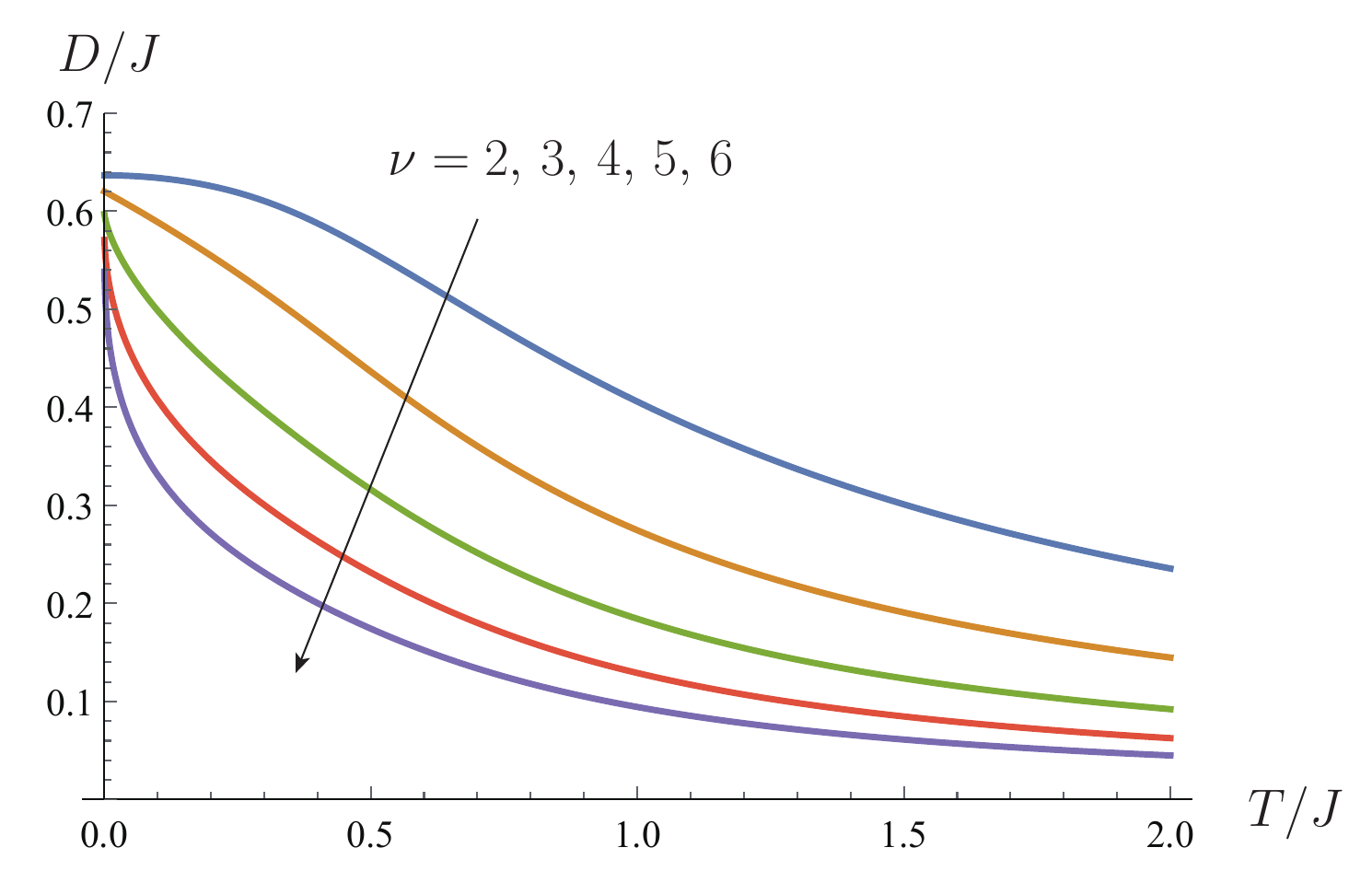}
\caption{The temperature dependence of the Drude weight $D$ for $2\le \nu \le
  6$ and $L=\infty$.}
\label{drude-T}
\end{figure}

The size-dependence of the Drude weight $D$ becomes more prominent with
increase of the interaction strength. However, $D$ eventually converges to
Zotos' results which are separately shown in Fig.~\ref{drude-T}.  Note that
the Drude weight $D$ converges to zero in the isotropic limit $\Delta=1$
\cite{Zotos}.
The dependence on the system size is more significant with decrease of the
temperature (or equivalently, with the decrease of the number of holes), which
follows from Fig~\ref{drude-nu}.  This strong size-dependence may explain that
extrapolations from various numerical methods such as numerical
diagonalizations and quantum Monte Carlo methods present contradictory results.
%
\section{Summary and Discussions}
%
In this paper, we have analyzed the spin transport properties of the spin-1/2
Heisenberg $XXZ$ chain for finite temperature.  We have constructed the $T$-
and $Y$-systems which are functional relations satisfied by the row-to-row
transfer matrices and certain combinations thereof ($Y$-functions).  This
enabled us to convert the problem of analyzing the energy spectrum into the
problem of examining the analytical properties of the
$Y$-functions. Consequently, the energy eigenvalues for arbitrary excited
states can be written as the solutions to non-linear integral equations
(NLIEs) satisfied by the $Y$-functions.  Combining this with the thermodynamic
Bethe ansatz (TBA) which determines the equilibrium state, we have succeeded
in describing the Drude weight in terms of specific $Y$-functions compatible
with the solutions to the TBA equations. Analyzing the NLIEs and TBA
equations, we have evaluated the large-size asymptotic behavior of the Drude
weight.  We found that the Drude weight sensitively depends on the system size,
i.e., the number of holes,
and shows very slow convergence in the thermodynamic limit $L\to\infty$ to the
results obtained by Zotos \cite{Zotos}. This sensitive and strong finite-size
dependence might explain the difficulties to evaluate the Drude weight by
extrapolations of numerical approaches.

There exist several natural extensions of our method.  The first one is an
extension to the case of general rational numbers $\nu=p/q$ where $p$ and $q$
are positive coprime integers satisfying $p\ge 2q$.  The $Y$-system
corresponding to \eqref{ysys} should be modified along the lines
  of \cite{KSS}.  We expect that the resultant formula of the Drude weight in
the thermodynamic limit will recover that derived by a different approach
\cite{UOKS}. In fact, the Drude weight at finite temperatures is considered to be
an everywhere discontinuous function of the anisotropy parameter $\Delta$: the
Drude weight exhibits a fractal dependence on $\Delta$ \cite{Pro11,PI,UOKS}.
For more quantitative and rigorous analysis of this intriguing behavior, a
formula describing the Drude weight for any irrational numbers $\nu$ is highly
desired.

Another simple extension is to consider the model with external magnetic
fields. This can be done by replacing the driving term $s_j(v)$
\eqref{driving} of the TBA equations \eqref{TBA} by
\begin{equation}
s_j(v)=(\delta_{j1}+\delta_{j2}\delta_{\nu2})s(v)
+ \frac{\nu}{2} \beta h(\delta_{j \nu-1}
-\delta_{j \nu}).
\end{equation}
In this case, the thermomagnetic effects such as the spin Seebeck effect
play a crucial role in the transport properties \cite{SK05,FIS,PZ}.

Finally, the effect of boundary conditions is also an interesting problem.
For the twisted boundary conditions with a twist angle $\alpha\in\mathbb{R}$,
we can derive a formula by just replacing $\phi\to \phi+\alpha$ in our
formulation. The analytic treatments, however, are more complicated, since all the
summands in \eqref{dressed-drude} contribute to the Drude weight, in contrast
to $\alpha=0$ where only two summands $j=\nu-1$, $j=\nu$ survive.

\section*{Acknowledgements}
The present work was partially supported by Grant-in-Aid for Scientific
Research (C) No. 16K05468 from Japan Society for the Promotion of Science and
by research unit FOR 2316 of DFG.
AK acknowledges interesting and helpful discussions with X.~Zotos.

\begin{appendix}

\section{Derivation of NLIEs}
%
Let us explain how to derive the NLIEs in section 4. Here we restrict
ourselves to the case $\nu> 3$. A similar treatment is also
possible for $\nu\le 3$.

In general, we solve the functional equations in multiplicative form by taking
the logarithmic derivative and apply the Fourier transform. In order to have a
Fourier representation with convergence in a sufficiently wide strip in the
complex plane we have to render  the functions analytic in such a
strip. This is achieved by suitable ``counter terms''.  For instance, the NLIE
for $j=1$ in \eqref{Y-sys} can be derived in the following manner. Due to
\eqref{yfunc1}, one finds that, in the physical strip $\Im v\in [-1,1]$, the
$Y$-function $Y_1(v)$ has zeros at $v=\zeta_2^k$ ($k=1,\dots,m_2$) and a 
{zero} of order $L$ at $v=0$. Define a modified function as
\begin{equation}
\widetilde{Y}_1(v)=\frac{Y_1(v)}{(\tnh \frac{\pi}{4}v)^L 
\prod_{k=1}^{m_2}\tnh{\frac{\pi}{4}}\left(v-\zeta_2^k \right)}.
\end{equation}
(Note that the second factor in the denominator is not needed for the $XY$ case
($\nu=2$), and one more additional factor $\prod_{k=1}^{m_3}\tnh\frac{\pi}{4}(v-\zeta_3^k)$
is needed for $\nu=3$).  Utilizing the identity
$\tnh\frac{\pi}{4}(v+\rmi)\tnh\frac{\pi}{4}(v-\rmi)=1$,
we rewrite the $Y$-system \eqref{ysys} as
\begin{equation}
\widetilde{Y}_1(v+\rmi)\widetilde{Y}_1(v-\rmi)=1+Y_{2}(v).
\label{Y1}
\end{equation}
The functions $\widetilde{Y}_1(v)$ and $1+Y_2(v)$ (see \eqref{yfunc2}) are
analytic and nonzero in the physical strip, and have constant asymptotics.
Taking the logarithmic derivative of both sides of \eqref{Y1}, applying the
Fourier transform
\begin{equation}
\widehat{f}(k)=\mathcal{F}[f(v)]:=\int_{\infty}^{\infty} f(v)\rme^{\rmi k v} \rmd v,
\end{equation}
and then shifting the integration contours of the left hand side of the resultant
equation, we have
\begin{equation}
2\ch k\, \mathcal{F}[\der_v\log \widetilde{Y}_1(v)]=\mathcal{F}[\der_v \log(1+Y_2(v))].
\end{equation}
Performing the inverse Fourier transform 
\begin{equation}
f(v)=\mathcal{F}^{-1}[\widehat{f}(k)]:=\frac{1}{2\pi}
\int_{-\infty}^{\infty} \widehat{f}(k)\rme^{-\rmi k v} \rmd v,
\end{equation} 
and using the convolution integral
\begin{equation}
\mathcal{F}^{-1}[f(k)g(k)]=f\ast g(v):=\int_{-\infty}^{\infty}f(v-x)g(x) \rmd x,
\end{equation}
we finally arrive at
\begin{equation}
\log Y_1(v)=L\log\tnh\frac{\pi}{4}v+\sum_{k=1}^{m_2}\log\tnh\frac{\pi}{4}(v-\zeta_2^k)
+s\ast \log(1+Y_2)(v),
\label{Y1-2}
\end{equation}
where
\begin{equation}
s(v):=\mathcal{F}^{-1}\left[\frac{1}{2\ch k}\right]=\frac{1}{4\ch\frac{\pi}{2} v}.
\end{equation}
The function $1+Y_2(v)$ appearing in the convolution in \eqref{Y1-2} has zeros
$\zeta_2^{k}+\rmi$ ($1\le k \le m_2$) which follows by \eqref{yfunc2}, and
hence we have
\begin{align}
\int_{\mathcal{L}_2} &s'(v-x) \log (1+Y_2(x)) \rmd x
=\int_{\mathcal{L}_2} s(v-x) \der_x \log (1+Y_2(x)) \rmd x \nn \\
&=-\sum_{k=1}^{m_2} \oint_{\zeta_2^k+\rmi}s(v-x)\der_x 
\log\left\{\sh\frac{\gamma}{2}(x-\zeta_2^k-\rmi) \right\}+s'*\log(1+Y_2)(v)\nn \\
&=-2\pi \rmi \sum_{k=1}^{m_2} s(v-\zeta_2^k-\rmi)+s'*\log(1+Y_2)(v) \nn \\
&=\der_v \log\tnh\frac{\pi}{4}(v-\zeta_2^k)+s'*\log(1+Y_2)(v),
\end{align}
where $\mathcal{L}_2$ is the contour modified such that it encircles
$\zeta_2^k+\rmi$ ($1\le k \le m_2$) (see Fig.~\ref{contour} in section 4).
The combination of this with \eqref{Y1-2} yields
\begin{equation}
\log Y_1(v)=L\log\tnh\frac{\pi}{4}v +s\underset{\mathcal{L}_2}{\ast}
\log(1+Y_2)(v).
\label{Y1-3}
\end{equation}
All the other NLIEs in \eqref{Y-sys} or \eqref{modified} can be derived in
completely the same way.

Finally let us derive the energy eigenvalue \eqref{spectrum} from the first
relation in \eqref{yfunc2}:
\begin{equation}
T_1(v+\rmi)T_1(v-\rmi)=\varphi(v+2\rmi)\varphi(v-2\rmi)(1+Y_1(v)).
\label{T1}
\end{equation}
Let us define
\begin{equation}
\widetilde{T}_1(v)=\frac{T_1(v)}{\varphi(v-\rmi)\prod_{k=1}^{m_1}
\tnh\frac{\pi}{4}(v-\zeta_1^k)}.
\end{equation}
(Note that, for $\nu=2$, one more additional factor
$\prod_{k=1}^{m_2}\tnh{\frac{\pi}{4}}(v-\zeta_2^k)$ is needed in the denominator.).  Then one
finds that $\widetilde{T}_1(v)$ is analytic and nonzero in {$\Im
v\in[-1-\epsilon_1, 1-\epsilon_2]$} 
where $\epsilon_1$ and $\epsilon_2$ are some small positive real
numbers,
and also $\widetilde{T}_1(v)$ has constant asymptotics. Eq.~\eqref{T1}
is then modified as
\begin{equation}
\widetilde{T}_1(v+\rmi)\widetilde{T}_1(v-\rmi)=
\frac{\varphi(v+2\rmi)}{\varphi(v)}(1+Y_1(v)).
\end{equation}
Applying a procedure similar to the derivation of the NLIEs \eqref{Y1-2} 
and \eqref{Y1-3}
\begin{align}
\der_v\log T_1(v-\rmi)&=\der_v\log \varphi(v-2\rmi)-2\pi \rmi {L} s\ast a(v)+
2\pi \rmi \sum_{k=1}^{m_1}s(v-\zeta_1^k)+s'\ast \log(1+Y_1)(v-\rmi) \nn \\
&=\der_v\log \varphi(v-2\rmi)-2\pi \rmi {L} s\ast a(v)+s'\astll \log(1+Y_1)(v-\rmi).
\end{align}
The application of the formula \eqref{baxter} into the above equation leads to
\eqref{spectrum}.
%
\section{Dressed function formalism}
%
Here we briefly summarize a technique to analyze the Drude weight $D$ given by
\eqref{drude-finite}. Let us consider the logarithmic derivative of
$Y$-functions:
\begin{align}
&\der \log(1+Y)=\frac{\der \log Y}{1+1/Y}, \nn \\
&\der^2\log(1+Y)=\frac{\der^2\log Y}{1+1/Y}+\frac{(\der\log Y)^2}{(1+1/Y)(1+Y)},
\end{align}
where $\der$ denotes the differential operator with respect to an arbitrary
parameter.  Applying this relations to {\eqref{modified}} (or \eqref{thermal}), one
obtains
\begin{align}
&(1+1/Y_j)A_j=a_j+\sum_{l=1}^\nu K_{jl}\astl A_l, \nn \\
&(1+1/Y_j)B_j=b_j+\sum_{l=1}^\nu K_{jl}\astl B_l,
\label{der}
\end{align}
where 
\begin{alignat}{3}
&A_j:=\der \log(1+Y_j), &\qquad & a_j:=\der p_j, \nn \\
&B_j:=\der^2 \log(1+Y_j), && b_j:=\der^2 p_j+\frac{(\der\log Y)^2}{1+Y}.
\end{alignat}
Multiplying the first (resp.~second) equation in \eqref{der} by $B_j$
(resp.~$A_j$), integrating over $v$ along the line $\mathcal{L}_j$ and summing
the resultant equation over $j$, we find the following is valid:
\begin{align}
\sum_{j=1}^{\nu} \int_{\mathcal{L}_j} \rmd v
&
\left\{\left(1+\frac{1}{Y_j(v)}\right)  A_j(v)  -a_j(v)\right\} B_j(v) \nn \\
&
\qquad=\sum_{j=1}^{\nu}\int_{\mathcal{L}_j} \rmd v
\left\{\left(1+\frac{1}{Y_j(v)}\right)B_j(v)-b_j(v)\right\} A_j(v).
\end{align}
Here we have used the relation  $K_{jl}(v)
  \left(=K_{lj}(v)\right)=K_{lj}(-v)$.  Thus we obtain
\begin{equation}
\sum_{j=1}^{\nu} \int_{\mathcal{L}_j} \rmd v\,
a_j(v) B_j(v) =\sum_{j=1}^{\nu}\int_{\mathcal{L}_j} \rmd v\,
b_j(v) A_j(v).
\end{equation}
Setting $A_j=\der_\psi \log(1+\Yt_j)$ and $B_j(v)=\der^2_\phi \log(1+\Yt_j)$
in \eqref{thermal}, and utilizing the above relation, we have
\begin{align}
\sum_{j=1}^\nu\int_{\mathcal{L}_j} 
\rmd v & \, \der_{\psi} p_j(v) \der_{\phi}^2 \log(1+\Yt_j(v)) \nn \\
&=\sum_{j=1}^\nu\int_{\mathcal{L}_j} \rmd v \,
\left\{\der_\phi^2 p_j(v)+
\frac{(\der_\phi \log \Yt_j(v))^2}{1+\Yt_j(v)}\right\}
\der_{\psi} \log(1+\Yt_j(v)).
\end{align}
The insertion of the relations $\der_{\psi} p_j(v)=s'(v{-i})\delta_{j1}$,
$\der^2_{\phi} p_j(v)=0$ yields
\begin{align}
{\sum_{j=1}^{1+\delta_{\nu 2}}
\int_{\mathcal{L}_j} }\rmd v  s'(v-\rmi) \der_{\phi}^2 &\log(1+\Yt_j(v)) \nn \\
&=
\sum_{j=1}^\nu \int_{\mathcal{L}_j} \rmd v
\frac{(\der_\phi \log \Yt_j(v))^2\der_{\psi} 
\log(1+\Yt_j(v))}{1+\Yt_j(v)}.
\end{align}
As a result, the Drude weight \eqref{drude-finite} is expressed as
\begin{equation}
D=\left.\lim_{L\to\infty}\frac{AL}{2\pi \rmi}\sum_{j=1}^\nu\int_{\mathcal{L}_j} \rmd v
\frac{(\der_\phi \log \Yt_j(v))^2\der_{\psi} 
\log(1+\Yt_j(v))}{1+\Yt_j(v)}\right|_{\phi=0,\psi=0}.
\label{dressed-drude}
\end{equation}

\end{appendix}


\end{document}